# Phonon mediated spin polarization in a one-dimensional Aubry-Andre-Harper chain

Moumita Mondal[1[0009-0008-7599-8799]] and Santanu K. Maiti[1[0000-0003-3979-8606]]

[1] *Physics and Applied Mathematics Unit, Indian Statistical Institute, 203 Barrackpore Trunk Road, Kolkata-700 108, India*
*mondalmoumita2k19@gmail.com, santanu.maiti@isical.ac.in*

**Abstract.** A comparative study of electronic transport and spin polarization between clean and Aubry-André-Harper chains in the presence of electron-electron (e-e) and electron-phonon (e-ph) interactions is presented. The entire system is simulated within a tight-binding framework based on the Hubbard-Holstein model. Transmission probability and spin polarization are evaluated using the Green's function method under the Hartree-Fock mean-field approximation through a self-consistent procedure. The transmission profile is found to be consistent with the band structure, which is also discussed. An overall enhancement of spin polarization induced by e-ph interaction is reported for the first time, to the best of our knowledge. Our analysis may be useful for studying controlled spin-selective electron transmission in the presence of e-ph coupling.

**Keywords:** Spin Selective Transmission, Spin Polarization, AAH Chain, Electron-Electron Interaction, Electron-Phonon Coupling, Hartree-Fock Mean-Field Theory, NEGF Technique

## 1 Introduction

A detailed understanding of electronic transport in low-dimensional systems has become essential for the fabrication of various nanoscale devices. This is the key reason that has captured researchers' attention toward exploring the current-carrying behavior of low-scale conductors [1,2]. As we know, transport in such systems is highly sensitive to electronic localization, scattering, presence of disorder, size of the systems, and many other factors. Moreover, as the quantum properties of materials become significant in this regime, particle-particle interactions [3–9] must also be taken into account to develop a comprehensive understanding. The motion of an electron near the Fermi surface is greatly influenced by its interactions with other electrons and with quantized lattice vibrations, known as phonons.

In this article, we consider a one-dimensional (1D) Aubry-André-Harper (AAH) chain [10-13] subject to electron-electron (e-e) and electron-phonon (e-ph) interactions, connected to source and drain leads. The interactions are incorporated following the Hubbard and Holstein prescriptions [8,9], where each lattice site is locally coupled to a dispersionless longitudinal optical phonon, and two electrons on the same site interact via a screened Coulomb repulsive force. All these phonons, having the same energy, are called Einstein phonons. A more realistic situation arises when, along with these elementary interactions, disorder effects are also introduced into the system, either in a correlated or an uncorrelated manner. Correlated disorder, which has advantages such as avoiding rigorous configuration averaging, is more suitable than random disorder. Among several possible forms of correlated disorder, the AAH model stands out because of its exceptional features, such as a nonzero transition point from a metallic to an insulating phase, a tunable phase factor, experimental realizability, and more [12,13,14]. To highlight the specific role of disorder, we compare our results with those of a disorder-free system.

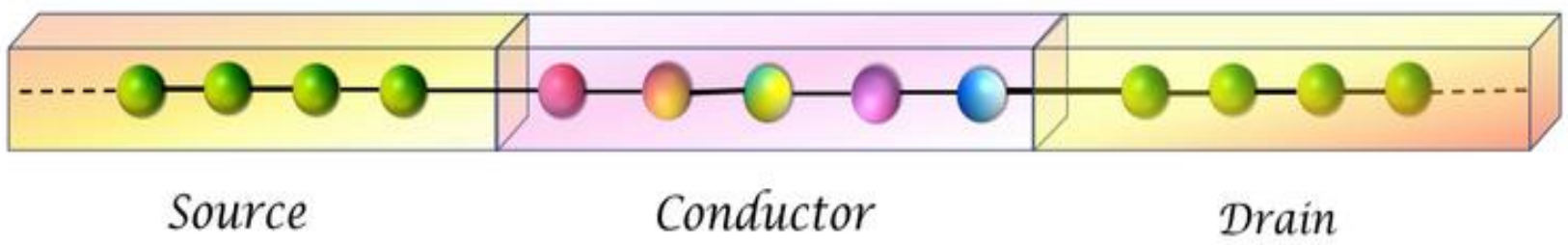

**Fig. 1.** Schematic diagram of lead-conductor nano-junction, where a 1D conductor, subject to electron-electron and electron-phonon interactions, is connected to two perfect semi-infinite 1D leads, the source and the drain.

The central focus of our work is to investigate the combined effects of AAH modulation, e-e interaction, and e-ph coupling on the transmission line shape and spin polarization (SP). To study the transport behavior, the chosen 1D chain is clamped between two contact leads (see Fig. 1), commonly referred to as source and drain leads, and the results are obtained using the well-known Green's function formalism [2,15]. The key aspects of our analysis are: (a) the appearance of antiferromagnetic ordering of magnetization in the half-filled Hubbard chain, which can be utilized to explore different anomalous signatures; (b) reduction of the effective bandwidths of the full energy window and the individual sub-bands; (c) atypical modification of different energy sub-bands due to the interplay between e-e and e-ph interactions in the presence of finite correlation among constituent atoms; and (d) observation of a high degree of spin polarization in the presence of disorder. Our analysis may provide a suitable route for achieving a high degree of spin polarization in different kinds of correlated, interacting quantum systems.

The rest of the paper is organized as follows. Section 2 presents the junction setup, tight-binding (TB) Hamiltonian, and the required mathematical tools for the calculations. Detailed descriptions of the results are given in Section 3. Finally, in Section 4, we conclude our findings.

# 2 Model and Theoretical Framework

## 2.1 Lead-Conductor Nanojunction

We begin with the schematic nanojunction shown in Fig. 1, where a one-dimensional (1D) non-magnetic conductor is connected to two 1D leads, referred to as the source and drain. Electron-electron (e-e) and electron-phonon (e-ph) interactions are considered within the conductor, while the side-attached leads are assumed to be free from these interactions. To model the source-conductor-drain nanojunction, we adopt a tight-binding framework. Since the electrodes are free from e-e and e-ph interactions, their TB Hamiltonians are straightforward and are not explicitly presented here. Instead, we focus on the conductor that bridges the two electrodes.

### 2.2 TB Hamiltonian of the conductor

For a conductor associated with e-e and e-ph interactions, the TB Hamiltonian can be written following the Hubbard-Holstein model as,

$$\mathcal{H}_{\text{ch}} = \sum_{i,\sigma} \varepsilon_{i\sigma} C_{i\sigma}^{+} C_{i\sigma} + t \sum_{i,j,\sigma} [C_{i\sigma}^{+} C_{j\sigma} + C_{j\sigma}^{+} C_{i\sigma}] + u \sum_{i,\sigma} C_{i\uparrow}^{+} C_{i\uparrow} C_{i\downarrow}^{+} C_{i\downarrow} \\ + \hbar\omega_0 \sum_{i} b_i^{+} b_i + g \sum_{i,\sigma} (b_i^{+} + b_i) C_{i\sigma}^{+} C_{i\sigma}, \qquad (1)$$

where $C_{i\sigma}^{+}$ and $C_{i\sigma}$ are the conventional Fermionic creation and annihilation operators at site *i* respectively, $t$ is the nearest-neighbor hopping strength, $u$ is the Hubbard interaction strength, $g$ denotes the e-ph coupling parameter, and $b_i^{+}$, $b_i$ are Bosonic operators associated with phonons.

The on-site potential of each site of the conductor is taken following the AAH model. For both up and down spin electrons, the site energies at any site *i* are given by,

$$\varepsilon_{i\uparrow} = \varepsilon_{i\downarrow} = W \cos(2\Pi b i + \phi_{\nu}), \qquad (2)$$

where $W$ denotes the AAH modulation strength, commonly referred to as the correlated disorder strength. The parameter *b* is an irrational number, which makes the site energies correlated, and $\phi_{\nu}$ is the AAH phase factor. To have a disorder-free system, we set $W$=0.

Through the Lang-Firsov unitary transformation [6, 9] followed by a zero-phonon averaging, the e-ph coupled Hamiltonian can be mapped onto an effective electronic Hamiltonian:

$$\mathcal{H}_{ch}^{eff} = \sum_{i,\sigma} \varepsilon_{i\sigma}^{eff} C_{i\sigma}^{+} C_{i\sigma} + t^{eff} \sum_{i,j,\sigma} [C_{i\sigma}^{+} C_{j\sigma} + C_{j\sigma}^{+} C_{i\sigma}] \\ + u^{eff} \sum_{i,\sigma} C_{i\uparrow}^{+} C_{i\uparrow} C_{i\downarrow}^{+} C_{i\downarrow}, \qquad (3)$$

where the effective parameters are defined as

$$\varepsilon_{i\sigma}^{eff} = \varepsilon_{i\sigma} - \frac{g^2}{\hbar\omega_0},\ t^{eff} = te^{-\left(\frac{g}{\hbar\omega_0}\right)^2},\ \text{and}\ u^{eff} = u - \frac{2g^2}{\hbar\omega_0}.$$

The last term of Eq. (3) is treated with the Hartree-Fock mean-field (MF) approximation [16,17], where the average occupation number of each site is evaluated self-consistently.

### 2.3 Green's Function Formalism

The spin-selective transmission probabilities are calculated using the well-known Green's function formalism [2, 15]. We need to define two types of Green's functions as

$$G_\sigma^R = [E\mathbb{I} - \mathcal{H}_{ch}^{eff} - \Sigma_S - \Sigma_D]^{-1},$$
$$G_\sigma^A = (G_\sigma^R)^+ \tag{4}$$

where $G_\sigma^R$ and $G_\sigma^A$ are the retarded and advanced Green's functions of the conductor, respectively. $\Sigma_{S/D}$ is the contact self-energy for the source (drain), which is also renormalized due to the e-ph coupling. For detailed calculation of self-energy expressions, see Ref. [2].

The transmission probability is computed from the relation,

$$T_\sigma(E) = Tr\,[\Gamma_S\, G_\sigma^R\, \Gamma_D\, G_\sigma^A], \tag{5}$$

where $\Gamma_{S/D} = -2\, Im\,(\Sigma_{S/D})$ is the broadening factor associated with the source (drain). Using the spin-dependent transmission probabilities, the spin polarization coefficient is evaluated as,

$$SP = \frac{T_\uparrow - T_\downarrow}{T_\uparrow + T_\downarrow} \times 100\ (\%) \tag{6}$$

where $T_\uparrow$ and $T_\downarrow$ are the transmission probabilities associated with up and down spin electrons, respectively. A maximum spin polarization is achieved when only one spin component transmits, while equal transmission of both spin components results in zero spin polarization.

## 3 Numerical Results and Discussion

We now turn our attention to the results obtained numerically based on the above theoretical discussions. Before starting a detailed analysis, we first mention the values of some input parameters that are kept constant throughout the study. For the side-attached leads, we choose $\epsilon_0 = 0$, and $t_0 = 2$. The conductor, a 30-site one-dimensional chain with hopping strength t = 1, is connected to the source and drain at the 1st and 30th sites, respectively. The coupling strengths between the chain and the leads are $\tau_s = \tau_d = 0.5$. The chain is taken to be half-filled, i.e., the total number of electrons is 30. The irrational number $b$ is chosen as the golden ratio, given by $b = (1+\sqrt{5})/2$, and the AAH phase factor is set to $\phi_\nu = 0$. All phonons are assumed to have energy $\hbar\omega_0 = 0.15$. Values of other physical parameters are mentioned at their respective places in the text. All energies are measured in units of eV, and the results are calculated at 0 K.

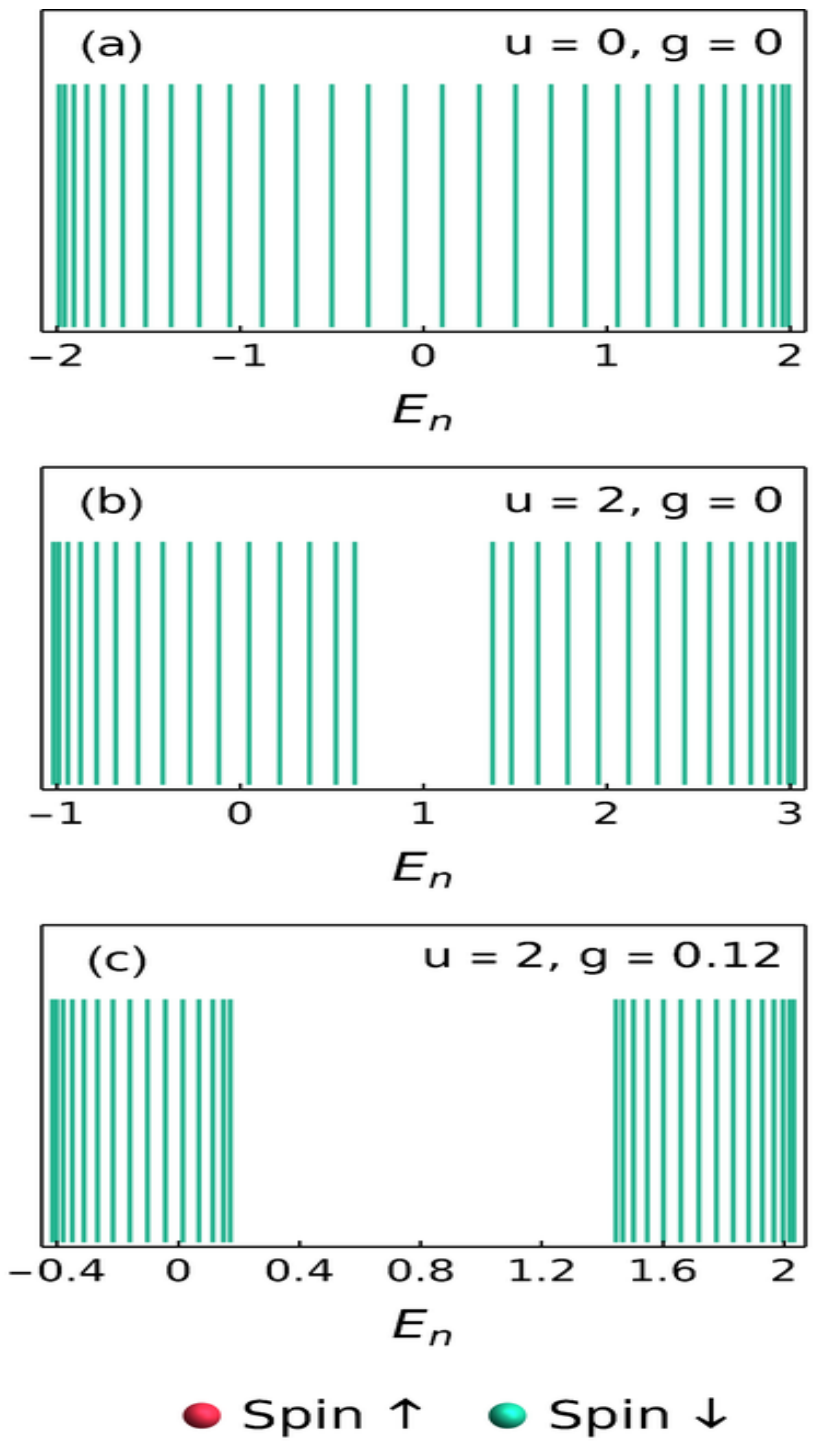

**Fig. 2.** Up and down spin energy eigenvalues for a 30-site 1D perfect chain for different values of $u$ and $g$. A vertical line is drawn in each energy eigenvalue for clearer visualization. The energy eigenvalues for the two spin cases match exactly.

Before analyzing the specific role of disorder, it is essential to first examine the characteristic features of the disorder-free conductor, i.e., when the disorder strength $W$=0. Figure 2 presents the energy eigenvalue spectrum for a perfect one-dimensional chain for different sets of electron-electron and electron-phonon coupling strengths, $u$ and $g$. In each case, a vertical line is drawn at every energy eigenvalue, with two different colors used to distinguish between the two spin states.

Figure 2(a) corresponds to the non-interacting situation, while Figures 2(b) and 2(c) represent the interacting cases. In the non-interacting limit (Fig. 2(a)), the energy eigenvalues are uniformly spaced, reflecting the typical linear dispersion of a perfect 1D chain. In contrast, Fig. 2(b), which includes electron-electron interaction, shows the formation of a band gap approximately at the center of the spectrum, accompanied by a noticeable shift of the entire band. Under the mean-field approximation, a perfect 1D Hubbard chain with an antiferromagnetic ground state behaves like a binary (AB-type) lattice. This effective binary structure induces a band gap whose magnitude depends on the interaction strength $u$. Additionally, the shift of the band arises from the modification of site energies due to the e-e interaction.

In Fig. 2(c), both electron-electron and electron-phonon interactions are taken into account. The presence of the e-ph coupling further enhances the band gap by narrowing the bandwidths on both sides of the spectrum. The additional shifting of the bands can again be attributed to the modification of the effective site energies, now incorporating contributions from phonons as well.

It is important to note that in all three cases, non-interacting, e-e interacting, and combined e-e and e-ph interacting, the eigenvalues for spin-up and spin-down electrons are exactly identical. This outcome stems from the structure of the Hamiltonians for the two spin channels. The underlying sub-lattice symmetry between the spin-up and spin-down Hamiltonians ensures that their eigenvalues remain identical, leading to a perfect superposition of the spin-up and spin-down bands.

From this observation, we can infer that under the present conditions, specifically in a disorder-free, perfect 1D conductor with the chosen interaction strengths, no spin-selective phenomena can be realized.

For a more detailed understanding, in Fig. 3 we present the spin-dependent transmission probabilities along with the corresponding spin polarization coefficient, considering different values of $u$ and $g$ under the half-filled band condition for a disorder-free conductor. The results are organized in three columns: the first column displays the transmission probabilities for up-spin electrons, while the second column shows those for down-spin electrons. The third column illustrates the spin polarization coefficients calculated from the respective transmission probabilities.

In accordance with the eigenvalue spectra discussed previously in Fig. 2, the transmission spectra exhibit distinct resonant peaks. Each of these resonant transmission peaks directly corresponds to an energy eigenvalue of the bridging conductor, thereby offering a direct mapping between the energy eigenvalues and the transmission characteristics. In other words, the transmission spectrum effectively mirrors the underlying energy spectrum of the conductor.

It is important to emphasize that the widths of these resonant transmission peaks are primarily governed by the strength of the coupling between the conductor and the electrodes. When the coupling is weak, the resonant peaks are extremely sharp and narrow, indicating well-defined energy levels. However, with increasing coupling strength, these peaks become progressively broader, reflecting stronger interactions between the conductor and the electrodes. The detailed influence of the coupling strength on the line shape of the transmission spectra has been extensively discussed in earlier works, and therefore, it is not elaborated here.

Another notable observation is that the transmission probabilities for up-spin and down-spin electrons are exactly identical across the entire energy range considered. Consequently, the spin polarization, which measures the difference between up and down spin transmissions, remains exactly zero throughout. Thus, a disorder-free conductor yields a vanishing spin polarization.

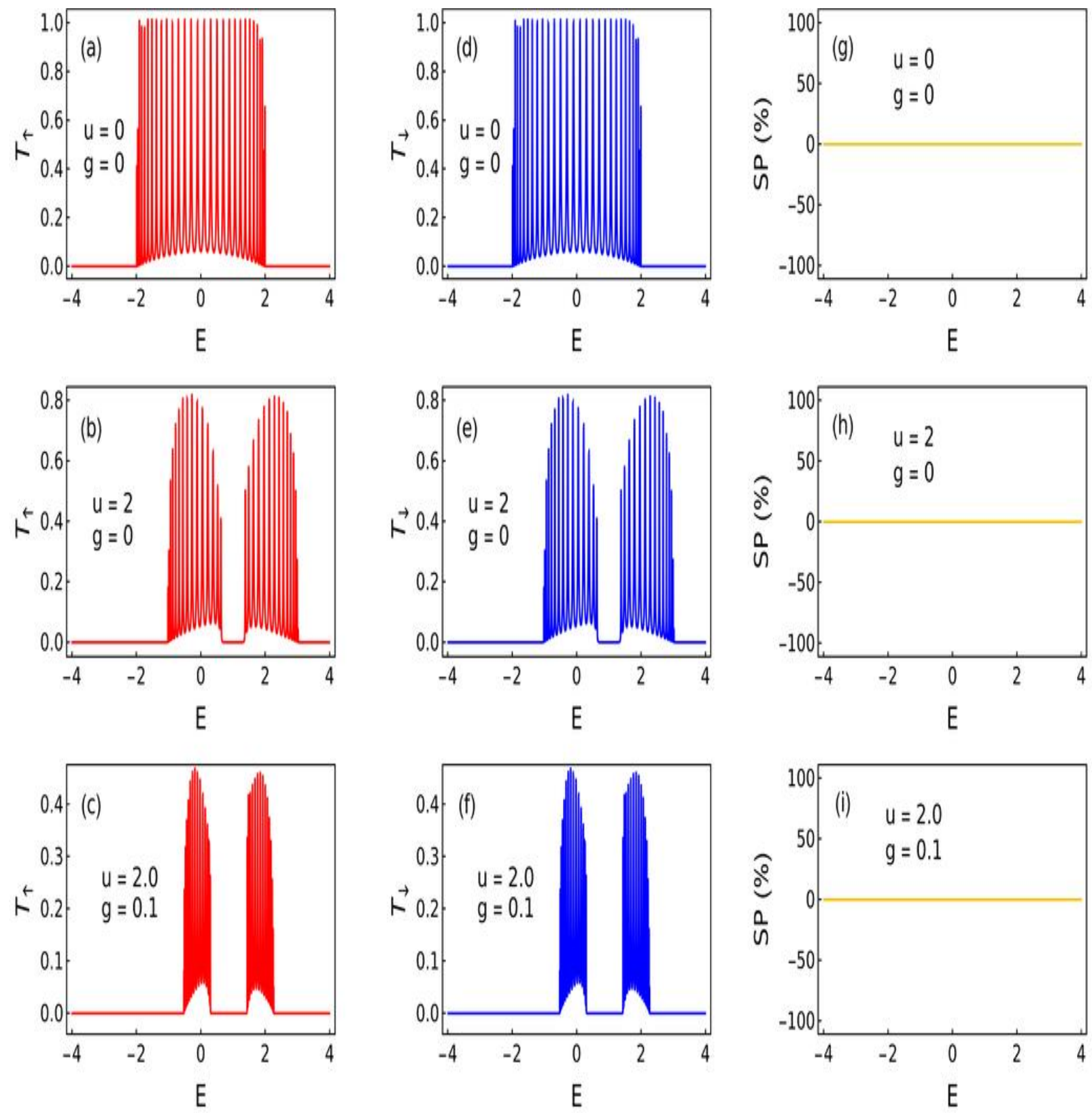

**Fig. 3.** Transmission probabilities, $T_{\uparrow}$ and $T_{\downarrow}$, for up and down spin electrons, together with the spin polarization coefficient as a function of energy, for a 30-site perfect conductor under different values of e-e and e-ph interaction strengths.

We now turn our attention to the correlated disordered system, which forms the core of our study. Our primary objective is to explore the intricate interplay among disorder, electron-electron interaction, and electron-phonon coupling. In line with the approach adopted for the disorder-free scenario, we begin our analysis by examining the energy eigenspectra of an isolated AAH chain under various input conditions characterized by different values of the interaction strengths $u$ and $g$. The corresponding results are presented in Fig. 4, where we consider a half-filled band configuration for a one-dimensional AAH chain consisting of 30 atomic sites.

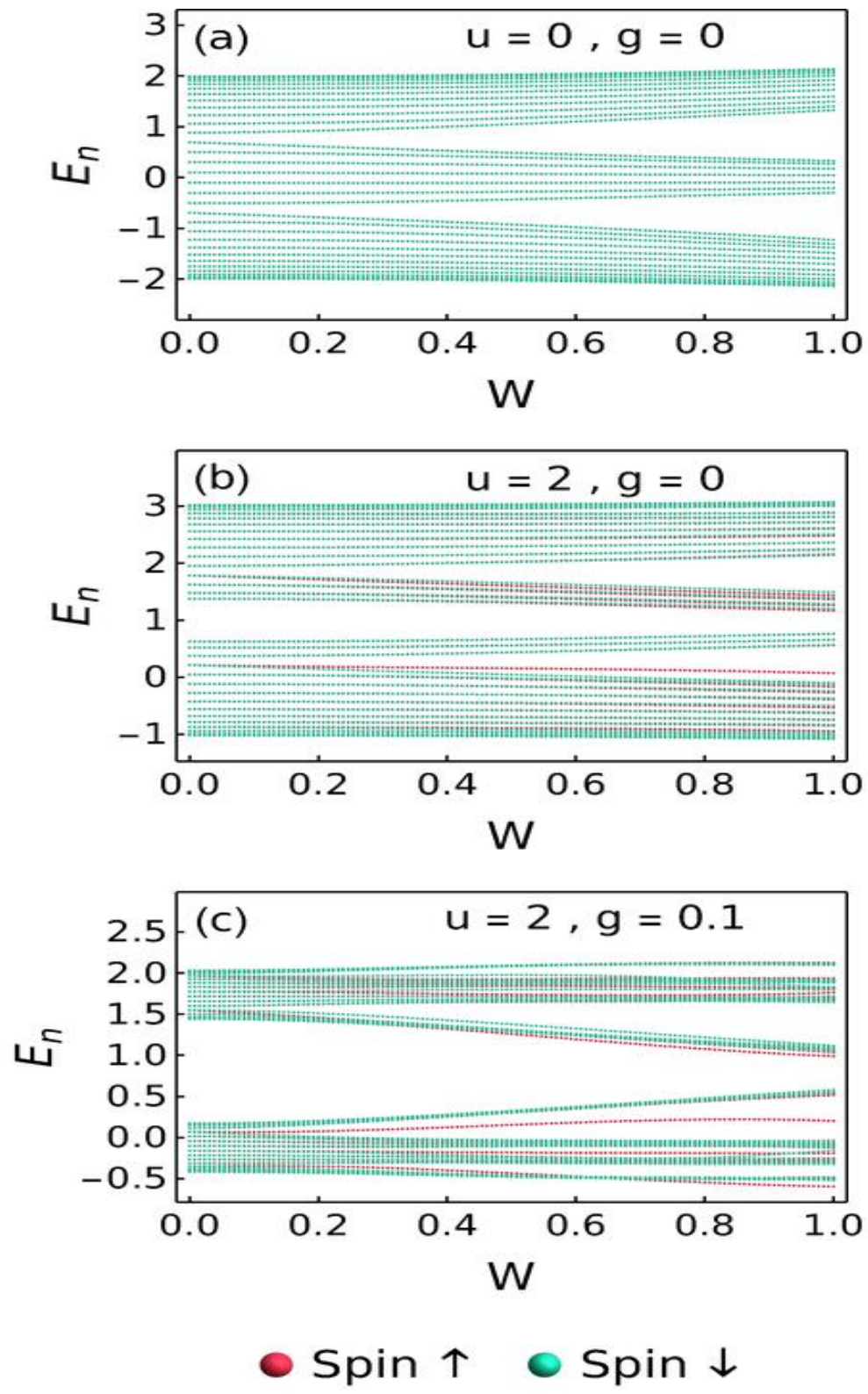

**Fig. 4.** Variation of energy eigenvalues for up and down spin electrons with disorder strength $W$ for a 30-site 1D AAH chain, considering three distinct sets of e-e and e-ph interaction strengths. The results correspond to the half-filled band case.

Figure 4(a) depicts the well-known energy band structure of a 1D AAH chain in the absence of any interaction. In this case, the energy levels associated with up-spin and down-spin electrons are exactly identical. This is expected, as the sub-Hamiltonians governing the up and down spin channels are identical when both e-e and e-ph interactions are absent. Consequently, no spin-dependent asymmetry arises.

The situation changes once the e-e interaction is introduced. As illustrated in Fig. 4(b), the presence of disorder in conjunction with e-e interaction breaks the spin symmetry. As a result, the up-spin and down-spin energy channels no longer coincide, leading to a visible mismatch between their corresponding energy eigenvalues. This spin-channel asymmetry originates from the spin-dependent effective potential created by the e-e interaction in a disordered background.

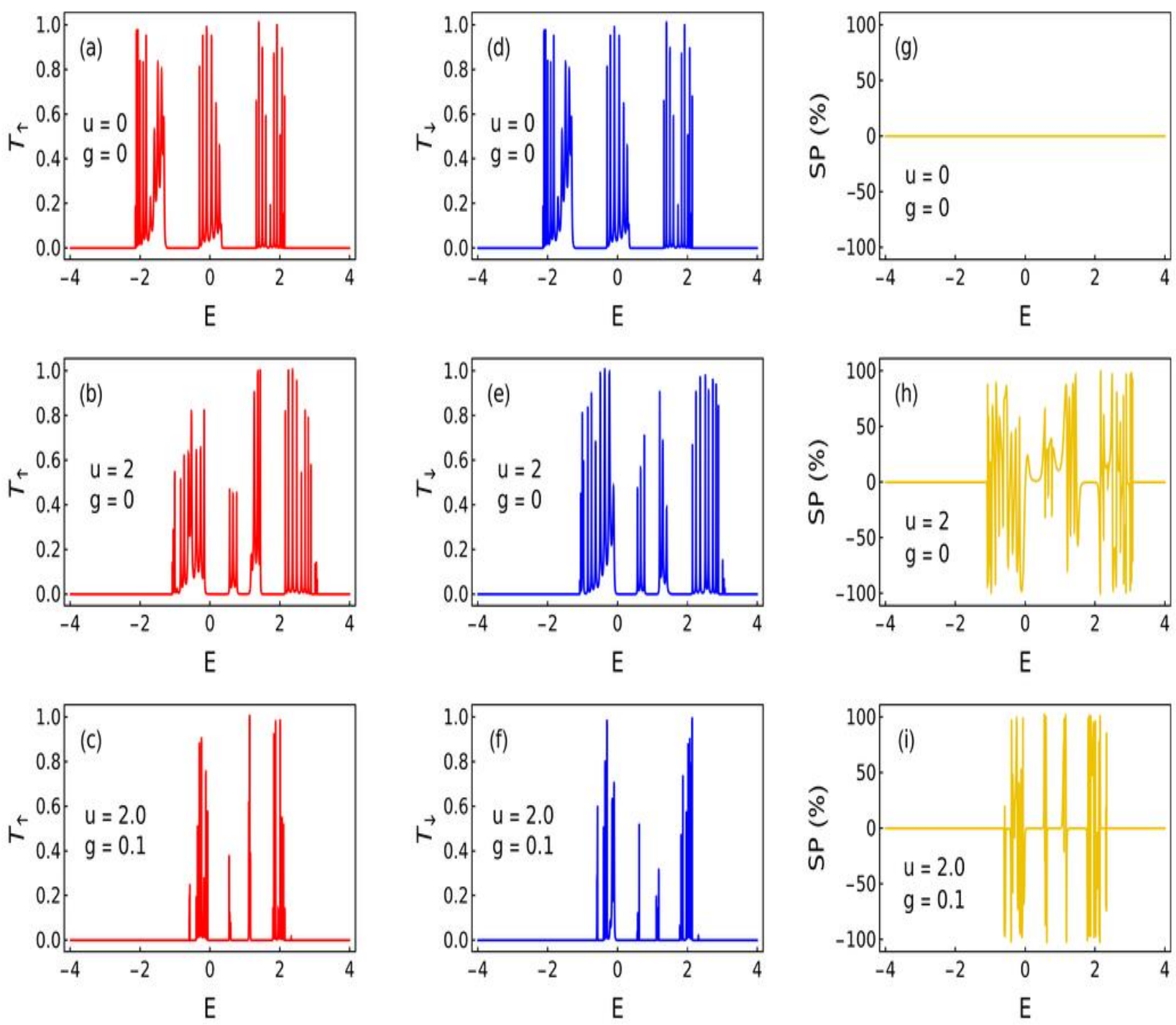

**Fig. 5.** Transmission probabilities, $T_\uparrow$ and $T_\downarrow$, for up and down spin electrons, together with the spin polarization coefficient as a function of energy, for a 30-site AAH chain under different values of e-e and e-ph interaction strengths. Here we choose $W$=1.

The disparity between the spin-resolved energy levels becomes even more pronounced when e-ph coupling is incorporated alongside the e-e interaction, as shown in Fig. 4(c). The additional presence of e-ph coupling further distorts the system, enhancing the symmetry breaking between the up-spin and down-spin sub-Hamiltonians. Furthermore, an overall narrowing of the energy bands is observed, which can be attributed to the renormalization effects induced by the electron-phonon interaction. The combined influence of e-e and e-ph interactions in a disordered setting thus leads to a significant restructuring of the energy spectra compared to the non-interacting case.

Figure 5 presents the characteristic behavior of spin-dependent transmission probabilities along with the spin polarization coefficient as a function of energy for the AAH chain, analyzed under different conditions involving e-e and e-ph interactions. The trends observed in these spectra are in close agreement with the energy eigenvalue distributions discussed earlier.

In the absence of both e-e and e-ph interactions, the transmission probabilities for up-spin and down-spin electrons are found to be exactly identical across the entire energy range. As a result, the spin polarization coefficient remains zero throughout, indicating a complete lack of spin-dependent transport under these conditions. How-ever, when e-e interaction is introduced (finite *u*), a distinct mismatch develops between the transmission profiles of up and down spin electrons. This mismatch leads to the emergence of a finite spin polarization at various energies. Notably, there are certain energy values where a remarkably high degree of spin polarization is achieved, highlighting the sensitivity of the system's spin transport properties to electron-electron interaction. The effect becomes even more pronounced upon the inclusion of e-ph coupling (finite *g*). To understand this behavior, it is important to consider the underlying mechanism: the mismatch between up and down spin channels is primarily governed by the asymmetry between the spin-resolved sub-Hamiltonians. This asymmetry is directly related to the ratio *u/t*. In the presence of e-ph coupling, both *u* and *t* experience a renormalization and generally decrease. However, the hopping integral *t* diminishes at a faster rate compared to *u*. As a result, the effective *u/t* ratio becomes larger when e-ph interaction is considered than when it is absent. A higher *u/t* ratio enhances the asymmetry between the spin channels, which in turn increases the mismatch between the up and down spin transmission probabilities. Consequently, this leads to a higher degree of spin polarization when electron-phonon interactions are present.

Thus, the combined influence of electron-electron and electron-phonon interactions plays a crucial role in engineering spin-polarized transport in the AAH chain, providing a potential route for tuning spintronic functionalities in low-dimensional quantum systems.

*Effect of temperature*: All the results discussed in this communication are obtained at absolute zero temperature. However, the effects of finite temperature can also be considered, which may yield interesting insights. At non-zero temperature, the effective hopping becomes temperature-dependent [5] due to the Bose-Einstein distribution of phonon occupation, thereby influencing spin polarization (SP) [18]. As temperature increases, thermal broadening causes greater overlap between the up- and down-spin transmission profiles, leading to a decrease in SP. Similar effects can arise from driven phonon populations or thermal pumping. Exploring these aspects presents a promising direction for future research.

*Possible tuning of SP*: Our findings indicate that a high degree of SP can be achieved in the presence of e-ph coupling. By varying the strength of this coupling, it may be possible to tune SP, which could be highly beneficial for designing efficient spintronic devices. Potential methods to modulate e-ph coupling include applying mechanical strain, inducing optical phonon excitation, substrate engineering, among others [19, 20]. Additionally, instead of focusing solely on a simple one-dimensional correlated system as in our current analysis, alternative materials such as two-dimensional moire superlattices or twisted geometries, where stronger e-ph interactions are expected [21, 22], could lead to even larger spin polarization.

*Experimental propositions*: For the sake of completeness in our discussion, it is important to consider the possible experimental realizations of our chosen system. With the advancements in nanotechnology today, such superlattices can be fabricated with relative ease. Typically, two laser beams having wave vectors $k_1$ and $k_2$ are directed from opposite directions to create a quasiperiodic potential, determined by the incommensurate parameter given by the ratio $k_1/k_2$. This setup allows for selective trapping of atoms [23, 24]. Additionally, more complex geometries can be realized by employing more intricate laser configurations, enabling the fabrication of various sophisticated structures.

## Closing Remarks

In this work, we have explored and demonstrated a new approach to enhance spin polarization in a simple one-dimensional chain, considering the combined effects of electron-electron and electron-phonon interactions within a two-terminal setup. Our analysis reveals that for an intermediate range of the electron-phonon coupling strength, the degree of spin polarization can increase with the enhancement of $g$. This finding suggests a new pathway to achieve a higher degree of spin polarization in such nanojunctions, a phenomenon that, to the best of our knowledge, has not been reported before.

However, it is important to note that this trend does not persist for all values of $g$. When $g$ becomes sufficiently large, the spin polarization starts to decrease with further increases in the coupling strength. This suppression can be attributed to the strong electron-phonon interaction effectively screening the Hubbard interaction, thereby diminishing the correlation effects that are crucial for maintaining spin polarization. Our study thus highlights a delicate balance between electron-electron and electron-phonon interactions in optimizing spintronic performance in low-dimensional systems.

## References

1. Imry, Y.: Introduction to mesoscopic physics. Oxford University Press, Oxford, (1997).
2. Datta, S.: Electronic transport in mesoscopic systems. Cambridge University Press, Cambridge (1997).
3. Essler, F.H.L.: The one-dimensional Hubbard model, Cambridge University Press, Cambridge (2005).
4. Lieb, E.H., Wu, F.Y.: The one-dimensional Hubbard model: A reminiscence. Physica A 321, 1 (2003).
5. Mahan, G.D.: Many particle physics. Plenum Press, New York (1990).
6. Mogulkoc, A., Modarresi, M., Kandemir, B.S., Roknabadi, M.R., Shahtahmasebi, N., Behdani, M.: The role of electron phonon interaction on the transport properties of Graphene based nano-devices. Physica B 446, 85 (2014).
7. Kolesnikov, D.V., Lobanov, D.A., Osipov, V.A.: The effect of electron-phonon interaction on the thermoelectric properties of defect zigzag nanoribbons. Solid State Communication 248, 83 (2016).
8. Stauber, T., Peres, N.: Effect of Holstein phonons on the electronic properties of graphene. J. Phys.: Condens. Matter. 20, 055002 (2008)

9. Lang, I., Firsov, Y.A.: Kinetic theory of semiconductor with low mobility. J. Exp. And Theor. Phys. 16, 1301 (1963).
10. Aubry, S., Andre, G.: Analyticity breaking and Anderson localization in incommensurate lattices. Ann. Israel Phys. Soc. 3, 133 (1980).
11. Harper, P.G.: Single Band Motion of Conduction Electrons in a Uniform Magnetic Field. Proc. Phys. Soc. A 68, 874 (1955).
12. Sil, S., Maiti, S.K., Chakrabarty, A.: Metal-Insulator Transition in an Aperiodic Ladder Network: An Exact Result. Phys. Rev. Lett. 101 (7), 076803 (2008).
13. Verbin, M., Zilberberg, O., Kraus, Y.E., Lahini, Y., Silberberg, Y.: Observation of Topological Phase Transitions in Photonic Quasicrystals. Phys. Rev. Lett. 110 (7), 076403 (2013).
14. Roy, S., Maiti, S.K.: Circular current in a one-dimensional Hubbard quasi-periodic Su-Schrieffer-Heeger ring. J. Phys.: Condens. Matter. 35, 355303 (2023).
15. Maiti, S.K.: Electron transport through mesoscopic ring. Physica E: Low-Dimens. Syst. Nanostruct. 36 (2), 199 (2007).
16. Maiti, S.K., Chakrabarti, A.: Magnetic response of interacting electrons in a fractal network: A mean-field approach. Phys. Rev. B 82, 184201 (2010).
17. Maiti, S.K.: Magnetic Response in Mesoscopic Hubbard Rings: a mean field study. Solid State Communication 150, 2212 (2016).
18. Fathizadeh, S.: Phonon assisted nearly spin current in DNA molecular chains: a multifractal analysis. Sci. Rep. 13, 21281 (2023).
19. Faugeras, C., Amado, M., Kossacki, P., Orlita, M., Sprinkle, M., Berger, C., de Heer, W.A., Potemski, M.: Tuning the electron-phonon coupling in multilayer graphene with magnetic fields. Phys. Rev. Lett. 103, 186803 (2009).
20. Ramar, M., Mariappan, M.S., Gunasekaran, S. et al.: Tuning electron-phonon coupling in Se-doped $Fe_2O_3$ for efficient photocatalysis: experimental and first-principles calculations. ACS Appl. Nano Mater. 7, 6935 (2024).
21. Chen, C., Nuckolls, K.P., Ding, S. et al.: Strong electron-phonon coupling in magic-angle twisted bilayer graphene. Nature 636, 342 (2024).
22. Gong, Z., Zhang, Q.Q., Mu, H.Y., An, X.T., Liu, J.J.: Spin polarized transport properties in magnetic moiré superlattices. Phys. Rev. B 109,045301 (2024).
23. Wright, E.M., Arlt, J., Dholakia, K.: Toroidal optical dipole traps for atomic Bose-Einstein condensates using Laguerre-Gaussian beams. Phys. Rev. A 63, 013608 (2000).
24. Timm, L., Ruffert, L.A., Weimer, H., Santos, L., Mehlstaubler, T.E.: Quantum nanofriction in trapped ion chains with a topological defect. Phys. Rev. Research 3, 043141 (2021).